\documentstyle[12pt,aasms4]{article}
\lefthead{LEAMON ET AL.}
\righthead{KINETIC DISSIPATION OF INTERPLANETARY TURBULENCE}

\begin{document}
\input{epsf.sty}

\title{Contribution of cyclotron-resonant damping to kinetic 
	dissipation of interplanetary turbulence}
\author{Robert J.\ Leamon, William H.\ Matthaeus, Charles W.\ Smith}
\affil{Bartol Research Institute, University of Delaware, Newark}
\authoraddr{R.~J.\ Leamon, W.~H.\ Matthaeus, C.~W.\ Smith,
	Bartol Research Institute, 
	University of Delaware, Newark, DE 19716. 
	(e-mail: [leamon, yswhm, chuck]@bartol.udel.edu)
}
\and
\author{Hung K.\ Wong}
\affil{Aurora Science, Inc., San Antonio, Texas}
\authoraddr{H.~K.\Wong, Aurora Science Inc.,
	4502 Centerview Drive, San Antonio, TX 87228.
	(e-mail: kit@cascade.gsfc.nasa.gov)
}

\begin{abstract}
We examine some implications of inertial range and dissipation range
correlation and spectral analyses extracted from 33 intervals of Wind 
magnetic field data.
When field polarity and signatures of
cross helicity and magnetic helicity are examined, most of the datasets 
suggest some role of cyclotron 
resonant dissipative processes involving thermal 
protons.
We postulate that an active spectral cascade into 
the dissipation range is balanced by a combination of cyclotron-resonant
and non-cyclotron-resonant kinetic dissipation mechanisms, of which 
only the former induces a magnetic helicity signature.
A rate balance theory, constrained by the data,
suggests that the ratio of the two mechanisms is of order unity.
While highly simplified, this approach appears to account for several 
observed features, and explains why complete cyclotron absorption, 
and the corresponding pure magnetic helicity signature,
is usually not observed.
\end{abstract}

\keywords{MHD --- turbulence}

\section{Introduction}

The solar wind plasma
displays many characteristics that can be reasonably well described by
a magnetohydrodynamic (MHD) fluid model 
(Tu \&\ Marsch 1995; Burlaga 1995),
  \nocite{TuMarschbook,Burlagabook}
including features that appear to be related to fluid
turbulence (Coleman 1968) 
  \nocite{Coleman68}
and MHD wave activity (Belcher \&\ Davis 1971). 
  \nocite{BelcherDavis71}
Within the context of a simple nonlinear MHD theory, one expects that a 
key feature is spectral cascade of energy from larger, energy-containing 
scales through an inertial range and ultimately into a dissipative range
(von K\'arm\'an and Howarth 1938; Batchelor 1970; 
 Mart\'{\i}nez {\it et al.}\ 1997).
  \nocite{KarmanHowarth38,BatchelorTHT,MartinezEA97}
An MHD description of the solar wind or other collisionless
plasmas in astrophysics is a drastic oversimplification,
and it is therefore significant that solar wind observations support 
the general picture of a turbulent MHD cascade from large to small 
scales (Jokipii 1973; Matthaeus \&\ Goldstein 1982; 
 Goldstein, Roberts \&\ Matthaeus 1995).
  \nocite{Jokipii73,MattGold82a,GoldsteinEA95a}
Indeed, unless turbulent transfer and decay are invoked,
it is difficult to explain the proton temperatures at 1~AU
(Coleman 1968), 
  \nocite{Coleman68}
the solar wind's general non-adiabatic temperature profile
(Freeman 1988; Richardson {\em et al.}\ 1995)
  \nocite{Freeman88,RichardsonEA95}
and the radial variation of the fluctuation levels 
(Zank, Matthaeus \&\ Smith 1996).
  \nocite{ZankEA96}

There is an appealing simplicity in an explanation of these features 
in terms of a cascade controlled by a larger scale energy-containing 
fluctuations 
(Matthaeus {\em et al.}\ 1994; Zank {\em et al.}\ 1996) 
  \nocite{MattEA94-ec,ZankEA96} 
and mediated by a self-similar inertial range 
(Tu, Pu \&\ Wei 1984; Verma, Roberts \&\ Goldstein 1995).
  \nocite{TuEA84,VermaEA95}
This model is in direct analogy with hydrodynamic turbulence
(Batchelor 1970).
  \nocite{BatchelorTHT} 
Nevertheless, there is an essential and theoretically challenging piece 
that is missing in this picture.
The cascade must terminate at small scales, possibly in a dissipation 
range in which processes occur that convert MHD fluctuation energy 
into plasma thermal energy.
The possible involvement of ion cyclotron activity in the
observed onset of steepening of solar wind magnetic field spectra
(at $\approx 1$ Hz) 
(Behannon 1976; Denskat, Beinroth \&\ Neubauer 1983)
  \nocite{Behannon75,DenskatEA83}
has been discussed for some time.
Generally, discussion of the collisionless damping of
interplanetary fluctuations has concentrated on Landau damping 
(Barnes 1966, 1979).
  \nocite{Barnes66,Barnes79a}
It is only recently 
(Goldstein, Roberts \&\ Fitch 1994; Leamon {\em et al.}\ 1998)
  \nocite{GoldsteinEA94a,LeamonEA98}
that attention has begun to focus on a broader framework for
explaining dissipation processes.
A theoretical perspective that invokes
kinetic theory to convert fluid scale energy to heat is needed, 
taking into account spectral transfer that continually resupplies 
the dissipation range through broad-band nonlinear couplings.
This paper provides a simple description of this process
based upon the assumption of a steady cascade, with the goal of 
explaining recently described features 
of the dissipation.

\section{MHD turbulence parameters}

It is useful to adopt a leading order description based upon 
incompressible turbulence, in view of the low level of interplanetary 
density fluctuations 
(Roberts {\em et al.}\ 1987), 
  \nocite{RobertsEA87b}
the observed density spectrum 
(Montgomery, Brown \&\ Matthaeus 1987), 
  \nocite{MontEA87}
and the low average turbulent Mach number 
(Matthaeus, Goldstein \&\ Roberts 1990).
  \nocite{MattEA90}
This perspective is also consistent with the persistence of the 
$k^{-5/3}$ signature of the Kolmogoroff cascade spectrum.
Neglecting small internal energy fluctuations, the turbulent
energy per unit mass, $E$, 
consists of
contributions from the turbulent (ion) velocity ${\bf v}$ and 
the fluctuating component of the magnetic field ${\bf b}$, 
scaled to Alfv\'en units.
For an appropriately defined ensemble average $\langle \dots \rangle$,
the contribution to the energy
from velocity fluctuations $E_v$ and
from magnetic fluctuations $E_b$ is
\begin{equation}
 E = E_v + E_b =  \frac{\langle |{\bf v}|^2 \rangle}{2} +
                  \frac{\langle |{\bf b}|^2 \rangle}{2}.
\label{eq:E}
\end{equation}

In its idealized definition, the turbulent energy includes contributions 
from all wavenumbers and frequencies. 
However, in some circumstances one might consider only contributions 
from certain scales, so that, for example, the spectral decomposition of 
magnetic energy,
$E_b = \int d^3 k E_b({\bf k})$
might include only a certain range of wavenumbers.
One might choose to look at the energy in a finite band of 
wavenumbers or frequencies, for example, when the physics of the 
inertial- or dissipation range is discussed.

Apart from energy, other quantities of importance for MHD turbulence 
are the magnetic helicity $H_m = \langle {\bf b} \cdot {\bf a} \rangle$, 
where ${\bf b} = \nabla \times {\bf a}$,
the cross helicity $H_c = \langle {\bf v} \cdot {\bf b} \rangle$, 
and the respective spectral decompositions 
(Matthaeus \&\ Goldstein 1982).
  \nocite{MattGold82a}
The amounts of cross helicity and magnetic helicity relative to the energy
are conveniently measured by the following dimensionless parameters.
The normalized cross helicity
\begin{equation}
 \sigma_c = \frac{E_+ - E_-}{E_+ + E_-},
\label{eq:sigmac}
\end{equation}
is defined in terms of the Els\"asser energies 
$E_\pm \equiv \langle |{\bf v} \pm {\bf b}|^2 \rangle$
(Marsch \&\ Mangeney 1987),
  \nocite{MarschMangeney87}
and lies between $-1$ and $+1$.
Normalized magnetic helicity
\begin{equation}
 \sigma_m = \frac{E_L - E_R}{E_L + E_R},
\label{eq:sigmam}
\end{equation}
is written here in terms of 
$E_L$, the magnetic energy in left-handed (positive helicity)
spatial structures, and 
$E_R$, the magnetic  energy in right-handed (negative helicity)
spatial structures. 
Note that $E_b = E_L + E_R$.
We use the following sense of circular polarization: 
right-handed means a 
sense of rotation from the $x$ direction towards the $y$ direction as 
one samples in the positive $z$ direction for a right-handed 
$({\bf x}, {\bf y}, {\bf z})$ coordinate system.
In terms of the integrated magnetic helicity spectrum,
$E^b_{L} = \frac{1}{2} \left( E_b + \int d^3k |{\bf k}| H_m({\bf k}) \right)$ 
and
$E^b_{R} = \frac{1}{2} \left( E_b - \int d^3k |{\bf k}| H_m({\bf k}) \right)$.

The magnetic helicity is important in the present context because
spatial handedness is related to resonance conditions with charged particles.
Cross helicity relates to the direction of propagation
of large amplitude Alfv\'en waves with respect to a uniform or 
slowly varying background magnetic field ${\bf B}_0$ 
(Belcher \&\ Davis 1971; Matthaeus \&\ Goldstein 1982).
  \nocite{BelcherDavis71,MattGold82a}
Both together determine the polarization of the waves in the plasma frame 
(Smith {\it et al.}\ 1984).
  \nocite{SmithEA84}

\section{Observations}

In a recent study, Leamon {\it et al.}\ (1998)
  \nocite{LeamonEA98}
described properties of the interplanetary dissipation range at 1~AU.
Their analysis included spectra and other parameters computed for
33 intervals of high time resolution (up to 22 vectors/s)
Wind magnetic field data, along with plasma data at a much lower 
sampling rate (either 46 or 92 seconds per measurement).
In this analysis, the magnetic field data provides
information about $E_b$, $E_L$ and $E_R$ in the dissipation range and
in the inertial range. 
For the samples in the Leamon {\it et al.}\ study, the inertial and
dissipation ranges were distinguished according to spectral slope.
The average inertial range spectral index corresponded to
a one-dimensional spectral law in good agreement with the 
Kolmogoroff value,
$E_b(k_r) \sim k_r^{-1.67}$, for radial wavenumber 
$k_r = 2\pi f / V_{SW}$,
with solar wind speed $V_{SW}$ and $f$ the spacecraft-frame frequency.
The dissipation range spectra were steeper, 
averaging $E_b(k_r) \sim k_r^{-3.01}$,
with a breakpoint between the 
two ranges at an average frequency of about $0.5$~Hz.

Leamon {\it et al.}\ noted that most of the intervals they examined 
showed a signature in the magnetic helicity at
dissipation range frequencies, as had been reported previously by
Goldstein {\it et al.}\ (1994).
  \nocite{GoldsteinEA94a}
In contrast, typical inertial range magnetic helicity spectra 
oscillate randomly as a function of frequency
(Matthaeus \&\ Goldstein 1982).
  \nocite{MattGold82a}
Leamon {\it et al.}\ found that as much as 90\%\ of the energy 
to be carried by waves propagating at highly oblique angles 
or quasi-two-dimensional turbulence rather than parallel-propagating 
Alfv\'en waves.
Nevertheless, in almost all of the intervals examined, the 
dissipation range $H_m$ were consistent with absorption of 
outward-propagating Alfv\'en waves by resonant coupling to 
thermal protons.

\placefigure{fig1}
Here we examine in greater detail the data underlying the latter 
conclusion.
In Fig.~1 we show the normalized cross helicity $\sigma_c$ computed 
from inertial range data, plotted versus
the normalized magnetic helicity $\sigma_m$ in the
dissipation range, for the 33 data intervals
previously analyzed.
$H_c$ can be computed only in the inertial range due to 
limited sampling rates for plasma data;
we use the inertial range $H_c$ as a proxy for the same quantity 
that is unmeasurable in the dissipation range.
In effect, we are assuming that the direction of propagation
of fluctuations in the dissipation range is
the same as the direction of propagation of fluctuations 
in the inertial range.

It is apparent from the data in Fig.~1 that most intervals
for which the mean magnetic field is outwardly directed have 
$\sigma_m >0$ and $\sigma_c < 0$.  
On the other hand, inwards directed ${\bf B}_0$ is associated with
$\sigma_m <0$ and $\sigma_c > 0$.
This implies a predominance of outward propagating waves.
One can readily see that this is consistent with cyclotron-resonant 
absorption of outward-propagating
fluctuations by thermal protons, as follows.
A proton moving outward along the magnetic field executes a 
left-handed helical trajectory.
Waves propagating outward at the Alfv\'en speed will overtake most 
thermal particles (at $\beta \approx 1$) and therefore, 
on average, the thermal protons will be in resonance with 
such waves that have a right-handed spatial handedness (negative $H_m$).
If the energy of these waves is assumed to be damped by the resonant 
protons, the energy that remains will preferentially reside in the 
undamped fluctuations, which have a left-handed structure and positive 
$H_m$ (see, for example, Moffatt (1978)).
  \nocite{Moffatttext}
Consequently, outward ${\bf B}_0$ should be associated with
$\sigma_c <0$ (outward propagation) and $\sigma_m > 0$.
Reversing the direction of ${\bf B}_0$ but maintaining the assumption
of outward propagating waves (now $\sigma_c >0$) produces the conclusion
that $\sigma_m < 0$ in the dissipation range by the same argument.

\section{Cascade and dissipation}

The above argument explains the clustering of the
observational points in the upper left and lower right quadrants.
However, there are questions that arise.
First, if kinetic processes are assumed to be very rapid,
why is the signature in the magnetic helicity not pure ($\pm 1$)
as one would expect for complete cyclotron absorption?
Second, how is the above argument modified if instead of pure 
cyclotron-resonant
absorption processes, there is also a contribution due to 
Landau resonance or nonresonant absorption?
Finally, since the observed cross helicities
are not ``pure,'' what is the effect of relaxing
the assumption of purely outward traveling Alfv\'en waves?

It turns out that these questions can be addressed, in at least a 
preliminary fashion, by postulating a cascade and associated
dissipation processes that are described by a set of
energy balance equations, as follows:
\begin{eqnarray}
 \frac {dE_L}{dt} & = &
 \frac {S}{2} - \gamma_0 E_L - \gamma_r E_L  \nonumber \\
 \frac{dE_R}{dt} & = & 
 \frac{S}{2}  - \gamma_0 E_R
\label{eq:rates-1}
\end{eqnarray}

The energies in left- and right-handed spatial structures 
are respectively designated as
$E_L$ and $E_R$ following our earlier discussion 
(in this case the integration over the spectrum now includes, 
by assumption, only the dissipation range).  
The rate of supply of energy (per unit mass) transferred into the 
dissipation range from the inertial range is designated by $S$. 
This supply rate is equally apportioned to $L$ and $R$ fluctuations 
since inertial range $H_m$ is random. 
We assume that the only external contribution to 
$dE_{L,R}/dt$ is due to the cascade term $S$, and that in the 
dissipation range there is no exchange between
$E_L$ and $E_R$, or exchange between kinetic and magnetic energies.
The quantity $\gamma_r$ represents a decay rate due to
cyclotron-resonant 
absorption by thermal protons, and it appears only in the 
$L$ equation under the assumption that fluctuations are outward 
propagating and ${\bf B}_0$ is inward. 
(This would also occur for inward propagation and outward ${\bf B}_0$.)
The remaining damping term, $\gamma_0$, appears in both $L$ and $R$ 
equations and represents decay processes that produce 
no signature in the magnetic helicity. 
Included in $\gamma_0$ are contributions from
Landau damping and other mechanisms
that do not involve cyclotron resonance, as well as
mechanisms that are fully nonresonant.

\section{Cyclotron-resonant and other forms 
of dissipation}

We can now proceed to estimate a typical relative strength of 
cyclotron-resonant
and non-cyclotron resonant processes.
Supposing the cascade is steady, so $dE_{L,R}/dt = 0$, and
we may equate the right hand sides of Eqs.~({\ref{eq:rates-1}). 
From the data, we take a typical value of magnetic helicity to be
$\sigma_m \approx -1/3$. 
This corresponds in Eq.~(\ref{eq:sigmam}) to $E_R = 2E_L$ in the 
dissipation range.
Then for consistency with Eqs.~(\ref{eq:rates-1})
we must have $\gamma_0 \approx \gamma_r$, 
indicating that cyclotron and non-cyclotron
absorption mechanisms are approximately of equal strength.

Since observed values of $H_c$ are not pure, 
the above argument should be refined to account for a distribution 
of propagation directions relative to the slower thermal protons.
Assume, then, that there is a probability $P(L)$ 
that fluctuations are propagating outward, which produces a resonance 
between left-handed structures and thermal protons, and implies 
the appearance of $\gamma_r$ in the $E_L$ equation.
Assigning the probability of inward propagation to be 
$P(R) = 1-P(L)$ implies that resonance between right-handed structures 
and thermal protons is weighted accordingly.
Therefore, the cascade balance equations become:
\begin{eqnarray}
 \frac{dE_L}{dt} & = & \frac{S}{2} - \gamma_0 E_L - P(L) \gamma_r E_L
							\nonumber \\
 \frac{dE_R}{dt} & = & \frac{S}{2} - \gamma_0 E_R  - P(R) \gamma_r E_R
\label{eq:rates-2}
\end{eqnarray}

According to the Els\"asser representation, fluctuations with energy
$E_-$ tend to propagate along the mean field ${\bf B}_0$ while fluctuations 
having energy $E_+$ tend to propagate antiparallel to ${\bf B}_0$.
We assume for simplicity that the probability that, at any location in the 
plasma, a typical thermal proton will ``see'' 
outward propagation is proportional to the average outward propagating 
energy. 
Thus,
\begin{equation}
 P(L) = \frac {E_-}{E_-+E_+} = \frac {1 + \sigma_c}{2}
\label{eq:PL}
\end{equation}
and therefore $P(R) = (1- \sigma_c)/2$.

With this interpretation, we can make use of the data in Fig.~1 to 
constrain our model and 
arrive at further insights about the dissipation processes.
We invoke the steady form of 
Eqs.~(\ref{eq:rates-2}) along with the definitions 
Eqs.~(\ref{eq:sigmac}), (\ref{eq:sigmam}) and~(\ref{eq:PL}), 
and assume that $\gamma_0$ and $\gamma_r$ are independent of
$\sigma_c$, $\sigma_m$ and other plasma turbulence parameters.
Eliminating $E_{L}$ and $E_{R}$, 
we conclude that
\begin{equation}
 \sigma_c = - \left( 1 + 2\frac {\gamma_0}{\gamma_r} \right) \sigma_m.
\label{eq:gen}
\end{equation}

The best-fit line forced through the origin is
$\sigma_c = -1.90 \sigma_m$, while 
the best-fit straight line through the data is
$\sigma_c = -1.80 \sigma_m + 0.10$.
Considering either 32 or 31 degrees of freedom accordingly, 
the reduced chi-squared values of the two fits are 
$\chi^2_r = 1.78$ and $1.55$.
Putting $\sigma_c = -1.90 \sigma_m$ in Eq.~(\ref{eq:gen}) implies 
that $\gamma_r = 2.22 \gamma_0$.
The other important consequence of Eq.~(\ref{eq:gen}) is that only when
$\gamma_0 = 0$ do pure Alfv\'en waves lead to purely helical states.

\section{Interpretation \&\ Discussion}

This preliminary attempt to understand the observed interplanetary 
dissipation range spectra, while clearly oversimplified, appears to 
contain some suggestive features.
We postulated an equation that balances cyclotron-resonant 
and non-cyclotron-resonant 
dissipation effects of kinetic origin with steady spectral transfer into 
the dissipation range due to MHD scale cascade processes.
This formal structure evidently has been able to
account for some of the observed properties of the distribution of
inertial range cross helicity and dissipation range magnetic helicities.
We note that several important approximations are implicit in our treatment.
For example, we do not account in any way for the energy in the 
velocity field, $E_v$ in the dissipation range. This might 
be an acceptable approximation in either of two cases: 
if $E_v$ and its dissipation rate are much smaller than $E_b$ and 
its cascade rate $S$; or 
if a proportionality or approximate equality exists between 
$E_v$ and $E_b$ in the dissipation range.
In the absence of a better theoretical guidance, as well as plasma data at 
the requisite frequencies, we prefer the latter explanation at present.
We can suppose, for example, that the ``Alfv\'en effect''
attempts to enforce near equipartition of velocity and magnetic fields,
or that the phenomenological dissipation rates we assumed in fact include 
some contributions that are mediated by couplings to the velocity field, 
which would be expected to be heavily damped at scales near the thermal 
proton gyroradius.

Motivated by the typical values of magnetic helicity in the dissipation 
range (Fig.~1), and assuming that 
all fluctuations propagate in one direction,
we estimated near equality of cyclotron
resonant contributions represented 
by $\gamma_r$ and other dissipative mechanisms
represented by $\gamma_0$.
Using the inertial range cross helicity to estimate the relative
likelihood of propagation direction produced
a refined estimate $\gamma_r \approx 2 \gamma_0$.

In this development we have been forced, due to limitations of 
spacecraft instrumentation, to use the inertial range
cross-helicity to compute a proxy for the propagation direction 
in the dissipation range.
The most notable limitation of this substitution derives from the 
possibility that preferential dissipation may lead to different 
cross-helicity values in the dissipation range, although we are not 
aware of any observational evidence for this.
Indeed, the connection between $\sigma_c$ and direction of
propagation may be complicated in the dissipation range 
by various kinetic wave modes such as whistlers.
On the other hand lower frequency observations of $\sigma_c$ 
(see Matthaeus \&\ Goldstein 1982) often indicate that a single
direction of propagation is dominant 
over several orders of magnitude of scale, which would tend to support
our extrapolation into the higher dissipation range frequencies.
In any case, the correlation evident in Fig.~1 appears encouraging
with regard to use of this proxy.

The present results provide some preliminary insights into the 
structure of the interplanetary dissipation range, but 
additional work needs to be done to better understand the physics of the 
kinetic dissipation mechanisms represented by $\gamma_0$ and $\gamma_r$.
For example, we expect on general grounds, that 
Landau and non-resonant processes 
should make a contribution to dissipation of three-dimensional,
MHD turbulent fluctuations, but an acceptable large amplitude theory
of such processes is not yet developed as far as we are aware.
Similarly, resonant dissipation, generally evaluated by linear Vlasov 
theory, requires improvement for the same reasons. 
In addition, linear theory makes no prediction about damping of 
purely transverse ``two-dimensional'' turbulence, which appears to be 
favored by MHD in the presence of a moderately strong mean magnetic field 
(Matthaeus {\it et al.}\ 1994).
  \nocite{MattEA94-ec}
In this regard one would expect that MHD turbulence would be accompanied by 
a turbulent induced electric field ${\bf E} = -{\bf v} \times {\bf b}$ 
that would produce stochastic acceleration of suprathermal
particles, and associated damping of the fluctuations.
Further developments in kinetic theory are required to describe 
dissipation that is nonlinear, anisotropic and driven by an MHD cascade.

\acknowledgments
This work is supported by
NASA grants NAG5-3026 and NAG5-7164, 
NASA subcontract NAG5-2848, 
and NSF grant ATM-9713595 to the Bartol Research Institute.
The participation of H.K.W.\ is supported by 
NASA contract NAS5-32484 and by a grant to the Goddard Space Flight
Center from the NASA Space Physics Theory Program.

\newpage
\figcaption[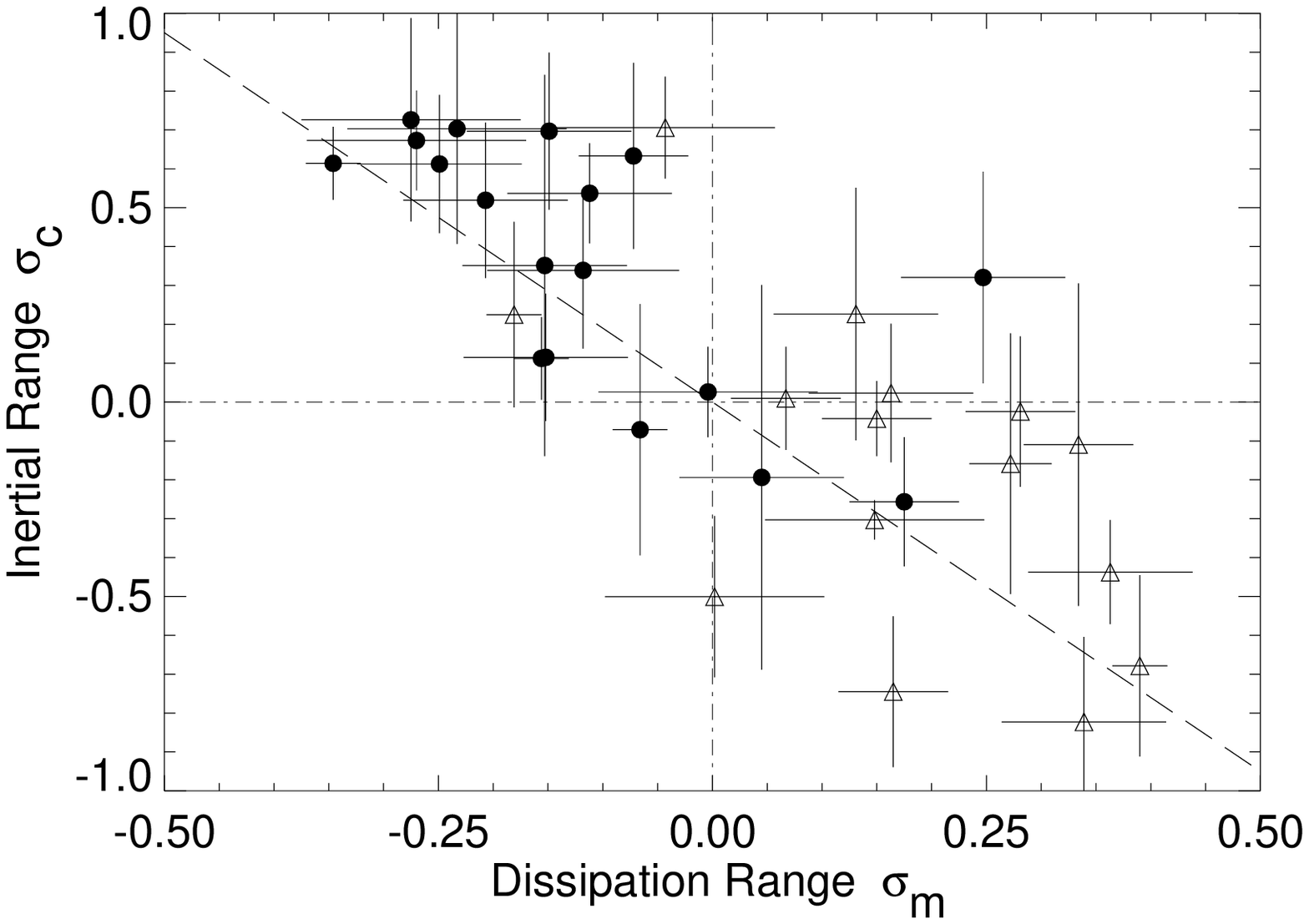]{Scatter plot, for 33 Wind data intervals,
	of the normalized cross helicity in the inertial range, $\sigma_c$,
	vs.\ the normalized magnetic helicity in the dissipation 
	range, $\sigma_m$.
	Triangles are intervals with outward directed mean magnetic 
	field, and bullets have inwards mean fields.
	The dashed line corresponds to the best-fit line through the
	origin, $\sigma_c = -1.90 \sigma_m$.
	\label{fig1}
}

\begin{figure}[p]
  \begin{center} 
  \leavevmode
  \epsfxsize=8.8cm 
  \epsfbox{leamon1.eps}
  \end{center}
\end{figure}


\newpage
\begin{thebibliography}{99}

\bibitem{Barnes66}
Barnes, A.\ 1966, {Phys.\ Fluids}, {\bf 9}, 1483


\bibitem{Barnes79a}
Barnes, A.\ 1979, in {\em Solar System Plasma Physics, vol. I},
  ed.\ E.~N.\ Parker, C.~F.\ Kennel, and L.~J.\ Lanzerotti,
  (Amsterdam: North-Holland), 251

\bibitem{BatchelorTHT}
Batchelor, G.~K.\ 1970, 
{\em The Theory of Homogeneous Turbulence},
(Cambridge: Cambridge University Press)

\bibitem{Behannon75}
Behannon, K.~W.\ 1976,
\newblock {\em Observations of the interplanetary magnetic field between 
	0.46 and 1~AU by the Mariner 10 spacecraft}, 
\newblock Ph.D.\ thesis, (Washington DC: Catholic Univ. of Am.)

\bibitem{BelcherDavis71}
Belcher, J.~W.\ and Davis, L., Jr.\ 1971, \jgr, {\bf 76}, 3534

\bibitem{Burlagabook}
Burlaga, L.\ 1995, {\em Interplanetary Magnetohydrodynamics}, 
(New York: Oxford University Press)

\bibitem{Coleman68}
Coleman, P.~J.\ 1968, \apj, {\bf 153}, 371

\bibitem{DenskatEA83}
Denskat, K.~U., Beinroth, H.~J., and Neubauer, F.~M.\ 1983,
{J.~Geophys.}, {\bf 54}, 60

\bibitem{Freeman88}
Freeman, J.~W.\ 1988, {Geophys.\ Res.\ Lett.}, {\bf 15}, 88

\bibitem{GoldsteinEA94a}
Goldstein, M.~L., Roberts, D.~A., and Fitch, C.~A.\ 1994,
\jgr, {\bf 99}, 11\,519

\bibitem{GoldsteinEA95a}
Goldstein, M.~L., Roberts, D.~A., and Matthaeus, W.~H.\ 1995,
\araa, {\bf 33}, 283

\bibitem{Jokipii73}
Jokipii, J.~R.\ 1973, \araa, {\bf 11}, 1

\bibitem{LeamonEA98}
Leamon, R.~J., Smith, C.~W., Ness, N.~F., Matthaeus, W.~H.,
and Wong, H.~K.\ 1998, \jgr, {\bf 103}, 4775

\bibitem{MarschMangeney87}
Marsch, E., and Mangeney, A.\ 1987, \jgr, {\bf 92}, 7363

 \bibitem{MartinezEA97}
 Mart\'{\i}nez, D.~O., Chen, S., Doolen, G.~D., Kraichnan, R.~H.,
 Wang, L.-P., and Zhou, Y.\ 1997, {J.\ Plasma Phys.}, {\bf 57}, 195

\bibitem{MattGold82a}
Matthaeus, W.~H., and Goldstein, M.~L.\ 1982, \jgr, {\bf 87}, 6011

\bibitem{MattEA90}
Matthaeus, W.~H., Goldstein, M.~L., and Roberts, D.~A.\ 1990,
\jgr, {\bf 95}, 20\,673

\bibitem{MattEA94-ec}
Matthaeus, W.~H., Oughton, S., Pontius, D.~H., and Zhou, Y.\ 1994,
\jgr, {\bf 99}, 19\,267

\bibitem{Moffatttext}
Moffatt, H.~K.\ 1978,
{\em Magnetic Field Generation in Electrically Conducting Fluids},
(New York: Cambridge University Press)

\bibitem{MontEA87}
Montgomery, D.~C., Brown, M.~R., and Matthaeus, W.~H.\ 1987,
\jgr, {\bf 92}, 282

\bibitem{RichardsonEA95}
Richardson, J.~D., Paularena, K.~I., Lazarus, A.~J., and Belcher, J.~W.\ 1995,
{Geophys.\ Res.\ Lett.}, {\bf 22}, 325

\bibitem{RobertsEA87b}
Roberts, D.~A., Goldstein, M.~L., Klein, L.~W., and Matthaeus, W.~H.\ 1987,
\jgr, {\bf 92}, 12\,023

\bibitem{SmithEA84}
Smith, C.~W., Goldstein, M.~L., Matthaeus, W.~H., and Vi\~{n}as, A.~F.\ 1984,
\jgr, {\bf 89}, 9159

\bibitem{TuMarschbook}
Tu, C.-Y., and Marsch, E.\ 1995,
{\em MHD\ Structures, Waves and Turbulence in the Solar Wind}.
(Dordrecht: Kluwer)

\bibitem{TuEA84}
Tu, C.-Y., Pu, Z.-Y., and Wei, F.-S.\ 1984,
\jgr, {\bf 89}, 9695

\bibitem{VermaEA95}
Verma, M.~K., Roberts, D.~A., and Goldstein, M.~L.\ 1995,
\jgr, {\bf 100}, 19\,839

\bibitem{KarmanHowarth38}
von K\'arm\'an, T.,  and Howarth, L.\ 1938, 
{Proc.\ Roy.\ Soc.\ London Ser.~A}, {\bf 164}, 192

\bibitem{ZankEA96}
Zank, G.~P., Matthaeus, W.~H., and Smith, C.~W.\ 1996,
\jgr, {\bf 101}, 17\,093

\end{thebibliography}
\end{document}